\documentclass[prb,twocolumn,showpacs,amsmath,amssymb]{revtex4-1}
\usepackage{amsmath,amssymb,amsfonts} 	
\usepackage{graphicx}
\usepackage{subfigure}

\renewcommand{\[}{\begin{equation}}
\renewcommand{\]}{\end{equation}}
\def\bea{\begin{eqnarray}}
\def\eea{\end{eqnarray}}
\def\nn{\nonumber\\}

\newcommand{\emi}[1]{{\rm e}^{-i #1}}
\newcommand{\ei}[1]{{\rm e}^{i #1}}
\newcommand{\B}{{\bf B}}

\newcommand{\intk}{\int_{\rm BZ}}

\newcommand{\M}{{\bf M}}
\newcommand{\MM}{{\mathfrak M}}
\newcommand{\CC}{{\mathfrak C}}
\newcommand{\MMM}{\mbox{\boldmath${\mathfrak M}$}}

\newcommand{\CP}{{\cal P}}

\newcommand{\CH}{{\cal H}}

\def\kk{{\bf k}}

\def\rr{{\bf r}}
\newcommand{\R}{{\bf R}}

\newcommand{\equ}[1]{Eq.~(\ref{#1})}
\newcommand{\eqs}[2]{Eqs.~(\ref{#1}) and (\ref{#2})}

\def\bra#1{\langle#1\vert}
\def\kket#1{\vert#1\rangle}
\def\ev#1{\langle#1\rangle}
\def\me#1#2#3{\langle#1| \, #2 \, |#3\rangle}
\def\runtime{(\the\time)\qquad\the\month/\the\day/\the\year}
\def\today
 {\count10=\year\advance\count10 by -2000 \number\day--\ifcase
  \month \or Jan\or Feb\or Mar\or Apr\or May\or Jun\or
             Jul\or Aug\or Sep\or Oct\or Nov\or Dec\fi--\number\count10}

\def\hour{\count10=\time\count11=\count10
\divide\count10 by 60 \count12=\count10
\multiply\count12 by 60 \advance\count11 by -\count12\count12=0
\number\count10 :\ifnum\count11 < 10 \number\count12\fi\number\count11}

\begin{document}
\title{Chern number and orbital magnetization in ribbons, polymers, and layered materials}

\author{Enrico Drigo$^1$ and Raffaele Resta$^{2,3}$}
\email{resta@iom.cnr.it}
\affiliation{$^1$Dipartimento di Fisica, Universit\`a di Trieste, Strada Costiera 11, 34151 Trieste, Italy}
\affiliation{$^2$Istituto Officina dei Materiali IOM-CNR, Strada Costiera 11, 34151 Trieste, Italy}
\affiliation{$^3$Donostia International Physics Center, 20018 San Sebasti{\'a}n, Spain}
\date{\today}

\begin{abstract}
The modern theory of orbital magnetization addresses crystalline materials at the noninteracting level: therein the observable is the $\kk$-space integral of a geometrical integrand. Alternatively, magnetization admits a local representation in $\rr$ space, i.e. a ``density'', which may address noncrystalline and/or inhomogeneous materials as well; the Chern number admits an analogous density. Here we provide the formulation for ribbons, polymers, nanowires, and layered materials, where both $\kk$-space and $\rr$-space integrations enter the definition of the two observables.
\end{abstract}

\maketitle 

\section{Introduction}

By definition, orbital magnetization $\M$ is the derivative of the macroscopic free-energy density with respect to the magnetic field (orbital term thereof, and with a minus sign). Customarily, the field adopted is ${\bf H}$; instead---because of the reasons well explained in Ref. \onlinecite{Griffiths-e}---first-principle theory adopts the more fundamental field ${\bf B}$. The modern theory of orbital magnetization dates since 2006; therein, the observable $\M$ is cast as the $\kk$-space  integral of a geometrical integrand.\cite{Vanderbilt} The expression addresses on the same ground trivial insulators, Chern insulators, and metals; it is also clear since then that $\M$ and the Chern invariant are closely related quantities on the theory side. In insulators the band spectrum is always gapped, but in the Chern case $\M$ depends on the value of the Fermi level $\mu$ in the gap: this behavior is consistent with the fact that, in a bounded sample, the spectrum is not gapped and the topologically protected edge states contribute to $\M$. 

In more recent years it has been shown that both the Chern invariant and $\M$ can also be defined for a bounded sample, where there is no $\kk$-vector to speak of: both 
observables can be computed by means of an $\rr$-space integration, where the integrand is to be regarded as the dual version of its $\kk$-space counterpart.\cite{rap146,rap148} While the $\kk$-space theory requires crystalline periodicity, the $\rr$-space approach is capable of dealing with strongly disordered cases and/or macroscopically inhomogeneous materials as well.
In this paper we are going to address the ``hermaphrodite'' cases,\cite{herma} i.e. those which require an integration over both $\rr$-space and $\kk$-space. The paradigmatic system in this class is a ribbon: a 2$d$ materials bounded in one Cartesian direction and lattice periodical in the other. The ribbon is also the simplest at the level of formulation and notations; the other hermaphrodite cases basically require to adopt the same logics within different notations. In this work we provide explicit formul\ae, thus extending the first-principle theory of both observables beyond their known formulation. In the ribbon case, we validate our expressions by means of model-Hamiltonian simulations.

The paper is organized as follows. In Sec. \ref{sec:ribbon} we show in detail the derivation of the ribbon magnetization formula; this also sets the logics to be adopted in the other hermaphrodite cases. The ribbon formula for the Chern number is derived as well. In Sec. \ref{sec:simulations} we provide a few test-case simulations based on the (by now famous) Haldane Hamiltonian.\cite{Haldane88} In the following Sec. \ref{sec:herma} we provide explicit formul\ae\ for the case of either a stereoeregular polymer or a nanowire (where only the normal component of $\M$ needs a nontrivial approach) and for the case of a layered material (where only the in-plane component of $\M$ was not accessible to the existing theory). Finally, in Sec. \ref{sec:conclusions} we draw some conclusions.

\section{Chern number and orbital magnetization in a ribbon} \label{sec:ribbon}

The 2$d$ orbital magnetization $M$ is a pseudoscalar with the dimensions of an orbital moment per unit area, while the Chern invariant is the (dimensionless) Chern number $C_1 \in \mathbb Z$. In the topological case $M$ as a function of $\mu$ in the gap is \[ M(\mu) = M(0) - \mu \frac{e}{hc} C_1,  \label{mudip} \] where the zero of $\mu$ is set at the top of the valence states; notice that the ribbon as a whole is gapless, but its bulk is insulating.
We address a ribbon of width $w$ in the $x$-direction, and lattice-periodical (``crystalline'') along $y$ with periodicity $a$ ("lattice constant''). The elementary definition of $M$ is given as the circulation of the microscopic orbital current per unit area: \[  M = \frac{1}{2 c w a} \int_{-\infty}^\infty dx \int_0^a dy \; x \; j_y^{(\rm micro)}(x,y) , \label{trivia} \] an expression dominated by boundary currents (we assume that the macroscopic current vanishes). As first proved in Ref. \onlinecite{rap148}, $\M$ is a local observable and admits a microscopic ``density'' called $\MM(\rr)$; in the case of a ribbon we have \[ M = \frac{1}{wa} \int_{-\infty}^\infty dx \int_0^a dy \; \MM(x,y) \label{smart} .\] The macroscopic average of $\MM(x,y)$ can be identified with (minus) the $B$-derivative of the free-energy density.  \eqs{trivia}{smart} provide an identical $M$ at any width $w$, but their {\it integrands} are very different. The transformation is similar in spirit to an integration by parts, and $\MM(x,y)$ is {\it not} a function of the microscopic orbital current.

What remains to be done is to express $M$ in terms of $1d$ Bloch orbitals, thus transforming the $y$-integral into a $k$-integral over the $1d$ Brillouin zone (BZ).  
We  expect that \eqs{trivia}{smart}  converge to the bulk $M$ value like $1/w$, but it will be shown that the present approach also allows to approach the large-$w$ limit in a more efficient way.

In all the hermaphrodite cases, the occupied orbitals obey the so-called open boundary conditions (OBCs) in some Cartesian direction(s), and Born-von-K\`arm\`an  periodic boundary conditions (PBCs) in some other(s). In order to address both cases, one needs a ``bridge'' providing a seamless path between the two frameworks. The key ingredient of this bridge must be the ground-state projector $\CP$ (a.k.a. the one-body density matrix), whose virtue is rooted in the ``nearsightedness'' principle, and which applies to $\CP$, but not to the individual eigenstates.\cite{Kohn96} Not surprisingly, $\CP$ is one essential ingredient of our formalism. 

 The microscopic magnetization density $\MM(\rr)$ may be cast in several equivalent forms; here---inspired by Ref. \onlinecite{Schulz13}---we adopt (in either 2$d$ or 3$d$) \bea \MMM(\rr)  &=& - \frac{i e}{2 \hbar c}  \me{\rr}{| \CH - \mu| \, [\rr , \CP] \times [\rr , \CP]}{\rr} \nn &=& \frac{e}{ \hbar c} \mbox{Im } \me{\rr}{| \CH - \mu| \, [\rr , \CP] \; [\rr , \CP]}{\rr}
 , \label{sch} \eea where $|\CH - \mu| = (\CH - \mu)({\cal I} - 2\CP)$, i.e. it is the operator which acts as $(\mu-\CH)$ over the occupied states, and as $(\CH - \mu)$ over the unoccupied ones. From \equ{sch} the $\mu$-derivative of $M$ in 2$d$ is \[ \frac{d }{d \mu}\MM(\rr) = \frac{e}{hc} 4\pi \, \mbox{Im }  \me{\rr}{|\CP \, [x , \CP] \; [y, \CP]}{\rr} = - \frac{e}{hc} \CC(\rr) , \label{raf} \] where $\CC(\rr)$ is a ``topological marker'' (a.k.a. Chern density), equivalent to the one defined in Ref. \onlinecite{rap148}; \equ{raf} is pespicuously consistent with \equ{mudip}.

\eqs{sch}{raf} are obviously well defined within OBCs, but it is not so obvious that they are well defined even in the crystalline case, where the Hamiltonian $\CH$ is lattice periodical and $\CP$ projects over a set of occupied Bloch orbitals. The multiplicative operator $\rr$ is a trivial one within OBCs, but is ``forbidden'' within PBCs: it maps a vector in the PBCs Hilbert space into something not belonging to the space.\cite{rap100} The commutators in \eqs{sch}{raf} effectively ``tame'' the nasty multiplicative $\rr$: this can be seen as follows. In a crystalline material the projector $\CP$ (as well as any other legitimate operator) is lattice-periodical: \[ \me{\rr}{\CP}{\rr'} = \me{\rr+\R}{\CP}{\rr'+\R}, \] where $\R$ is a lattice vector. It is immediate to verify that $i [\rr, \CP]$ is indeed a legitimate, lattice periodical, Hermitian operator.

 The Bloch orbitals of a ribbon are $\kket{\psi_{jk}} = \ei{ky} \kket{u_{jk}}$, with $\ev{x,y|u_{jk}}$ square-integrable along $x$ and periodical along $y$; we normalize them as \[  \int_{-\infty}^\infty dx \int_0^a dy \; \ev{x,y|u_{jk}}\ev{u_{j'k}|x,y} = \delta_{jj'} . \] The $\kket{u_{jk}}$ are eigenstates of $\CH_k = \emi{ky} \CH \ei{ky}$ with eigenvalues $\epsilon_{jk}$. Within these notations, the ground-state projector per spin channel is, in the Schr\"odinger representation: \bea \me{\rr}{\CP}{\rr'} &=& \frac{a}{2\pi}\intk dk \; \ei{ky} \me{\rr}{\CP_k}{\rr'} \emi{ky'}, \label{intk} \\
 \CP_k &=& \sum_{\epsilon_{jk} \leq \mu} \kket{u_{jk}} \bra{u_{jk}}.  \eea 
The integrand in \equ{intk} is a periodic function of $k$ with period $2\pi/a$, ergo the BZ integral of its $k$-derivative vanishes: \bea 0 &=& \frac{a}{2\pi} \intk dk \;  \frac{d}{dk} (\ei{k(y-y')} \me{\rr}{\CP_k}{\rr'}) \\  &=& i(y-y') \me{\rr}{\CP}{\rr'} + \frac{a}{2\pi} \intk dk \; \ei{ky} \me{\rr}{\CP'_k}{\rr'}\emi{ky'} , \nonumber \eea where $\CP'_k$ is the $k$-derivative of $\CP_k $. In compact operator notations this reads \[ i[y,\CP] = - \frac{a}{2\pi} \intk dk \; \ei{ky} \CP'_k \emi{ky} , \]  while the other commutator is identically written as \[ [x,\CP] = \frac{a}{2\pi} \intk dk \; \ei{ky} [x,\CP_k] \emi{ky'} ; \]
we also cast the Hamiltonian in a similar form: \[ \CH = \frac{a}{2\pi} \intk dk \; \ei{ky} \CH_k \emi{ky} ,\] where the integrand is actually $k$-independent. We have by now all the ingredients needed to evaluate the triple product in \equ{sch}, ergo in \equ{smart}. The triple $k$-integration contracts to one (details are given in the Appendix) and we get \bea \MM(\rr) &=& \frac{e}{\hbar c} \mbox{Im} \left(\frac{ia}{2\pi} \intk \frac{dk}{2\pi} 
\me{\rr}{| \CH_k - \mu| \, [x, \CP_k] \, P'_k}{\rr} \right) \nn &=& \frac{ea}{hc} \mbox{Re}
\intk \frac{dk}{2\pi} \me{\rr}{| \CH_k - \mu| \, [x, \CP_k] \, P'_k}{\rr}. \label{intkk} \eea This is a periodic function of $y$; we take the trace over the $1d$ cell to obtain the (microscopic) linear magnetization density (as a function of the bounded coordinate) \[ {\cal M}(x) = \frac{e}{hc} \mbox{Re}
\intk \frac{dk}{2\pi}  \; \mbox{Tr$_y$} \{\;| \CH_k - \mu| \, [x, \CP_k] \, P'_k\; \}. \label{major} \] \[ M = \frac{1}{w} \int_{-\infty}^{\infty} dx \; {\cal M}(x) \label{minor} .\]
\equ{major} is one of the major results of this work: it obviously yields $M$ via \equ{minor}, but---as shown below---can be used in a more efficient way by averaging  it in the central region of the ribbon (not on the whole ribbon).

The analogous formulae for the topological marker and for the Chern number are \[ {\cal C}(x) = - 4 \pi\, \mbox{Re}
\intk \frac{dk}{2\pi}  \; \mbox{Tr$_y$} \{\CP_k  \, [x, \CP_k] \, P'_k\; \}. \label{c1} \] \[ C_1 = \frac{1}{w} \int_{-\infty}^{\infty} dx \; {\cal C}(x) \label{c2} .\]

\begin{figure}[t]
\centering
\includegraphics[width=0.99\linewidth]{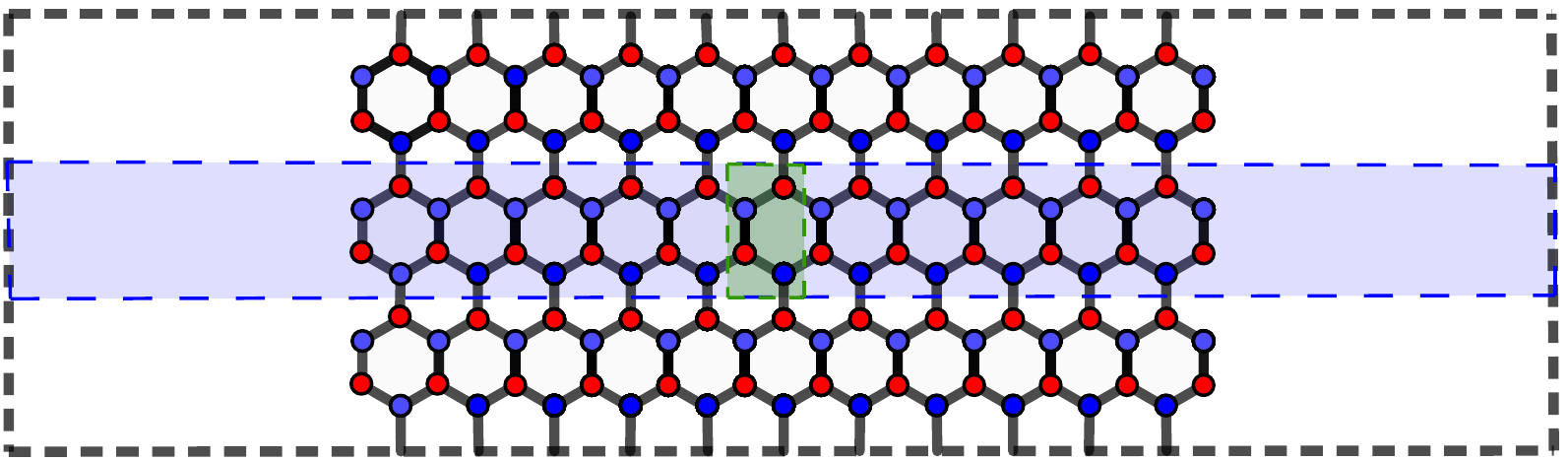} 
\caption{(color online). A typical armchair Haldanium ribbon; the linear cell (46 sites in the figure) and the central cell (4 sites) are shown. We performed simulation up to $\simeq 100$  sites in the linear cell.
}
\label{fig:ribbon} \end{figure}

Owing to gauge invariance, the $k$-derivative of $\CP$ can be safely evaluated by numerical differentiation. In a tight-binding implementation (as the one shown below) the evaluation of the trace amounts to perform multiplications of small matrices. In a first-principle implementation it would perhaps be more convenient to write the expression in terms of the $\kket{u_{jk}}$ in the ``Hamiltonian gauge'',\cite{Vanderbilt} i.e. using \equ{intkk} as it is.

\section{Simulations for an ``Haldanium'' ribbon} \label{sec:simulations}

\begin{figure}[t]
\centering
\includegraphics[width=0.99\linewidth]{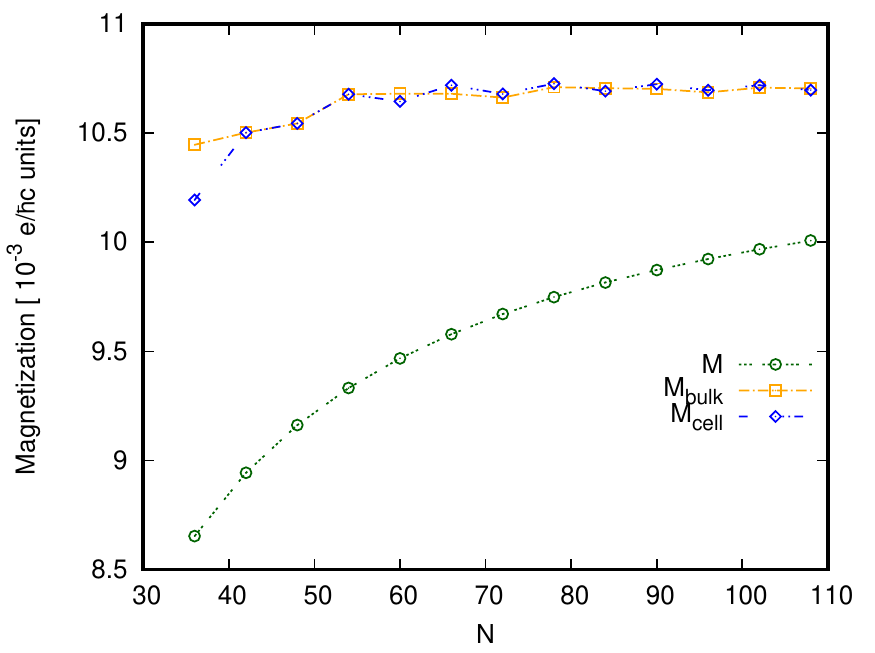} 
\caption{(color online). Orbital magnetization in the nontopological case  as a function of the ribbon width. The symbols $M$, $M_{\rm bulk}$, and $M_{\rm cell}$ are defined in the text. Units of $e/(\hbar c)$}
\label{fig:n-topo} \end{figure}

\begin{figure}[b]
\centering
\includegraphics[width=0.99\linewidth]{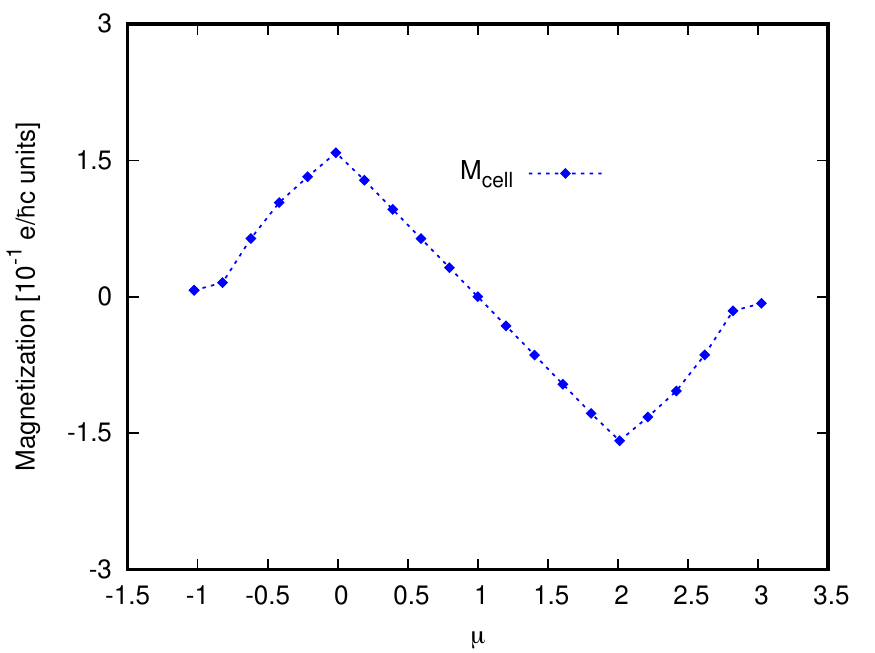} 
\caption{Central-cell orbital magnetization $M_{\rm cell}$ in the topological case  as a function of the Fermi level $\mu$ in units of $e/(\hbar c)$. After \equ{mudip} the $\mu$-derivative of $M$ is $-1/(2\pi) \simeq -0.159$ in the plot units.}
\label{fig:mu} \end{figure}

\begin{figure}[t]
\centering
\includegraphics[width=0.99\linewidth]{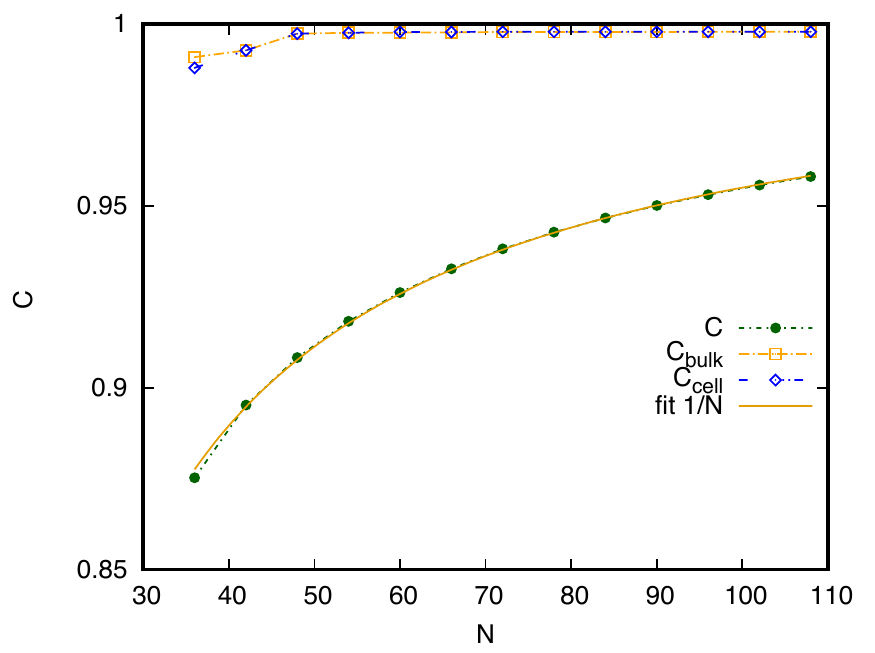} 
\caption{Convergence of the Chern number with the ribbon width. $C$, $C_{\rm bulk}$, and  $C_{\rm cell}$ are defined in the text, in analogy to Fig. \ref{fig:n-topo}}
\label{fig:chern} \end{figure}

The paradigmatic model for validating results of the present kind is the one proposed by Haldane in 1988.\cite{Haldane88} It is a tight-binding $2d$ Hamiltonian on a honeycomb lattice with onsite energies $\pm \Delta$, first neighbor hopping $t_1$, and second neighbor hopping $t_2 = |t_2| \ei{\phi}$, which provides time-reversal symmetry breaking. The model is insulating at half filling and metallic at any other filling; according to the parameter choice the insulator can be either trivial or topological (Chern number $ \pm 1$). The model has been previously used to demonstrate the locality of $\M$, and implemented for bounded samples within OBCs in order to address insulators (both trivial and topological)\cite{rap148,rap151} and metals.\cite{rap150} We are going to benchmark the present hermaphrodite results by implementing the model Hamiltonian on a periodic ``Haldanium'' ribbon, as the one shown in Fig. \ref{fig:ribbon}.

We evaluate $C_1$ and $M$ at finite widths, and we study their convergence as a function of the number of sites, using three different expressions. In the figures the symbol $M$ results from \equ{minor}, $M_{\rm bulk}$ results from averaging over 1/2 of the sites in the central region, $M_{\rm cell}$ results from averaging over the central cell (four sites); similar symbols are adopted for $C_1$. In the topological case the ribbon is gapless; we therefore adopt a ``smearing'' technique, common to many metallic simulations. In order to present homogeneous results, we adopt the smearing even in the topologically trivial case. 

As a prototype of nontopological Haldanium we adopt the parameters $t_1=1$, $t_2=1/3$, $\Delta=1.5$, $\phi = \pi/4$ at half filling; this choice also allows benchmarking to Ref. \onlinecite{rap150}, where some simulations adopt the same Haldanium parameters.
As a prototypical topological ($C_1 =1$) case we adopt a nonpolar case: $t_1=1$, $t_2=1/3$, $\Delta=0$, $\phi = \pi/2$; the simulations have been performed at various $\mu$ values.

We start displaying the nontopological case in Fig. \ref{fig:n-topo}. As expected---and consistently with \equ{trivia}---the integral over the whole ribbon converges only like $1/w$; the other curves converge instead much faster, owing to the quasi-exponential decay of the projector $P$ in the present insulating case. Next we switch to our topological case study; it has been proved that even in this case the projector $P$ has a quasi-exponential decay,\cite{Thonhauser06} and in fact the plot (not reproduced here) is qualitatively quite similar to Fig. \ref{fig:n-topo} for any $\mu$ value in the bulk gap. Next we show, for the same topological case, the value of $M_{\rm cell}$, as defined above, when the Fermi level $\mu$ is varied across the gap in \equ{major}. The perspicuous linear behavior is due to the filling of the topologically protected boundary states.

Finally in Fig. \ref{fig:chern} we show the convergence of our topological marker ${\cal C}(x)$, \equ{c1}, to the Chern number $C_1$, where $C$, $C_{\rm bulk}$, and  $C_{\rm cell}$ are defined analogously as for the magnetization plots. Here again the convergence is $1/w$ when ${\cal C}(x)$ is averaged over the whole ribbon, as in \equ{c2}, but it is exponential when the average is performed over an inner sample region.

\section{Polymers and layered materials} \label{sec:herma}

As said above, there is a family of hermaphrodite cases: (i) 2$d$ materials bounded in one Cartesian direction and lattice periodical in the other (ribbons, dealt with above); (ii) $3d$ materials bounded in 2 directions and lattice-periodical in the third (stereoregular polymers and nanowires, where only the normal $\M$ component is problematic); (iii) $3d$ materials bounded in one direction and lattice-periodical in the remaining two (where the in-plane component of $\M$ is problematic). Above we have discussed and demonstrated---via tight-binding simulations---the test case of a ribbon. The corresponding formul\ae\ for cases (ii) and (iii) above are reported in the following.

\subsection*{Magnetization of polymers and nanowires}

We deal here with a T-breaking quasi-1$d$ system, periodic along $z$ with period $a$. The Bloch orbitals are $\kket{\psi_{jk}}= \ei{kz}\kket{u_{jk}}$, normalized as
\[  \int_{-\infty}^\infty dx \int_{-\infty}^\infty dy \int_0^a dz \; \ev{\rr|u_{jk}}\ev{u_{j'k}|\rr} = \delta_{jj'} , \] and the ground-state projector is  (per spin channel): \bea \me{\rr}{\CP}{\rr'} &=& \frac{a}{2\pi}\intk dk \; \ei{kz} \me{\rr}{\CP_k}{\rr'} \emi{kz'}, \label{intk2} \\
 \CP_k &=& \sum_{\epsilon_{jk} \leq \mu} \kket{u_{jk}} \bra{u_{jk}}.   \eea 
Since the system is microscopic in the $(x,y)$ plane, the intensive quantity of interest $\M$ is defined as (minus) the ${\bf B}$-derivative of the free-energy {\it per unit length} (although the system is actually 3-dimensional).

The $z$-component af $M$ is simply proportional to the orbital moment per unit length: \bea M_z &=& -\frac{e}{2ac}\int_{-\infty}^\infty dx \int_{-\infty}^\infty dy \int_0^a dz \nn &\times& [\;  x \; j_y^{(\rm micro)}(\rr) -  y \; j_x^{(\rm micro)}(\rr)  ] ;\eea this is well defined since the system is bounded in the $(x,y)$ directions. The normal components requires instead to be addressed via the modern theory. From the main text it follows that \bea M_x &=&  \frac{1}{a} \int_{-\infty}^\infty dx \int_{-\infty}^\infty dy \int_0^a dz \; \MM_x(\rr) \; , \nn  \MM_x(\rr) &=& \frac{e}{ \hbar c} \mbox{Im } \me{\rr}{| \CH - \mu| \, [y , \CP] \; [z, \CP]}{\rr} . \eea The commutator $[z, \CP]$ is then transformed as in the main text. After contracting the $k$-integrals (see the Appendix) we get, similarly to the ribbon formula in Sec. \ref{sec:ribbon}: \[ \MM_x(\rr) = \frac{ea}{hc} \mbox{Re}
\intk \frac{dk}{2\pi} \me{\rr}{| \CH_k - \mu| \, [x, \CP_k] \, P'_k}{\rr} . \] 

\subsection*{Magnetization of lattice-periodical slabs}

We consider a 3$d$ system which is bounded in the $z$ direction and lattice periodical in the $(x,y)$ coordinates. The Bloch orbitals are $\kket{\psi_{jk}}= \ei{(k_x x + k_y y)}\kket{u_{jk}}$, normalized as
\[  \int_{-\infty}^\infty dz \int_{\rm BZ} d\kk\; \ev{\rr|u_{j\kk}}\ev{u_{j'\kk}|\rr} = \delta_{jj'} , \] where $\kk$ is the 2$d$ Bloch vector and BZ the relative Brillouin zone.
The ground-state projector is  (per spin channel): \bea \me{\rr}{\CP}{\rr'} &=& \frac{A_{\rm c}}{(2\pi)^2}\intk d\kk \; \ei{\kk \cdot \rr} \me{\rr}{\CP_k}{\rr'} \emi{\kk\cdot \rr'}, \label{intk3} \\ \CP_\kk &=& \sum_{\epsilon_{j\kk} \leq \mu} \kket{u_{j\kk}} \bra{u_{j\kk}},  \eea where only the $(x,y)$ components of $\rr$ enter the products $\kk \cdot \rr$. The  $\kket{u_{j\kk}}$ are eigenstates of $\CH_\kk= \emi{\kk\cdot \rr} \CH \ei{\kk\cdot \rr}$; notice that $\CH_\kk$ is a 3$d$ Hamiltonian, and $\kk$ is a 2$d$ Bloch vector.

The intensive quantity $\M$ of interest is the magnetization per unit area. The $z$ component of $\M$ can be derived from the standard modern theory of orbital magnetization, as shown in the original literature. Here we address the in-plane component of $\M$; if $A_{\rm c}$ is the area of the 2$d$ unit cell,  then  \bea M_x &=&  \frac{1}{A_{\rm c}} \int_{-\infty}^\infty dz  \int_{A_{\rm c}} dx\,dy \; \MM_x(\rr) \; , \nn  \MM_x(\rr) &=& \frac{e}{ \hbar c} \mbox{Im } \me{\rr}{| \CH - \mu| \, [y , \CP] \; [z, \CP]}{\rr} . \eea Here again we transform only the commutator 
$[y,\CP]$: \[ i [y,\CP] = - A_{\rm c} \intk \frac{d\kk}{(2\pi)^2} \;\ei{\kk\cdot \rr} (\partial_{k_y} \CP_\kk) \emi{\kk\cdot \rr} , \] while the other two entries in the matrix element are 

 \bea \CH &=& A_{\rm c} \intk \frac{d\kk}{(2\pi)^2} \;\ei{\kk\cdot \rr}  \CH_\kk \emi{\kk\cdot \rr} \nn {[z,{\CP}]} &=&  A_{\rm c} \intk \frac{d\kk}{(2\pi)^2} \;\ei{\kk\cdot \rr} [z,  \CP_\kk] \emi{\kk\cdot \rr} . \eea
After contracting the three $\kk$-integrals (see the Appendix) we get, in analogy to the ribbon case,
 \[ \MM_x(\rr) = \frac{e A_{\rm c}}{ \hbar c} \mbox{Re } \me{\rr}{| \CH_\kk - \mu| \, (\partial_{k_y} \CP_\kk) \; [z, \CP_\kk]}{\rr} . \]

\section{Conclusions} \label{sec:conclusions}

We have shown how to extend the theory of orbital magnetization beyond the two cases dealt so far in the literature: periodic crystalline materials, where $\M$ is the reciprocal space integral of a geometrical integrand;\cite{Vanderbilt} and bounded samples (possibly noncrystalline), where the magnetization density has a well defined expression in $\rr$-space.\cite{rap148,rap150} Similarly, the Chern number enjoyed a known dual picture.\cite{rap146} Here we have completed the theory of orbital magnetization, providing explicit formul\ae\ for all the cases which require integration over both reciprocal space and coordinate space. 

We have also provided a formulation for the Chern number $C_1$ in a ribbon geometry; the study of its convergence as a function of the ribbon width $w$ yields some important comments. Our formula converges like $1/w$ when integrated over the whole ribbon, and instead exponentially when integrated in a more efficient way (see text). When an unbounded crystalline sample is considered,  $C_1$ is computed as a $\kk$-integral on a 2$d$ BZ: in this case even a coarse $\kk$-mesh provides the converged result.\cite{Fukui05} If instead we address a flake (a sample bounded in both Cartesian directions), the integral over the whole flake is zero: the boundary therefore yields an extraordinary negative contribution,\cite{rap146} and the topological marker $\CC(\rr)$ {\it does not} average to one over a line. The boundary acts as a ``reservoir'': the marker may equal one in the bulk only if the boundary provides a negative compensating contribution.

The fundamental reason for the difference between unbounded samples and bounded samples is that the trace of the commutator of two finite-size matrices is zero, while the commutator of two unbounded operators may have a nonzero diagonal. In our ribbon case the 
operator is bounded in the $x$ direction and unbounded in the $y$ one: the nontrivial message from Fig. \ref{fig:chern} is that---at variance with the flake case---there are no extraordinary boundary contributions. A ``reservoir'' is not needed: the average of ${\cal C}(x)$ over the whole ribbon converges indeed to $C_1$, although slowly. 

Last but not least, the case of a finite $\B$ field is worth a comment. For bulk materials a macroscopic field is incompatible with PBCs (except in commensurate cases); the modern theory only addresses spontaneous magnetization. Instead, in all the hermaphodite cases the adoption of the appropriate Landau gauge yields a periodical Hamiltonian. The present formulation can therefore by applied in principle even to cases where a finite $\B$ field is present. Care has to be taken, though, because of the ubiquitous presence of Landau levels. The problem is highly nonanalytical at $\B=0$, and the density of states changes qualitatively in an abrupt way as soon as $\B\neq 0$ is set.\cite{rap150}
\bigskip

\section*{Acnowledgments}
Useful discussions with A. Marrazzo and R. Bianco are gratefully acknowledged. Work supported by the ONR Grant No. No. N00014-17-1-2803.

\section*{Appendix: Products of lattice-periodical operators}

We are going to make use of a simple  lemma, about the integral of a plane wave $\ei{ky}$ times a periodic 
function $f(y)$: \[ \int_{-\infty}^\infty dy \; \ei{ky} f(y) = \frac{2\pi}{a} \delta(k) \int_0^a dy' \; f(y') . \label{lemma} \] Any ribbon-periodic operators ${\cal A}$ and ${\cal B}$ in the Schr\"odinger representation can be written as: 
\bea \me{\rr}{{\cal A}}{\rr'} &=& \frac{a}{2\pi}  \intk dk \; \ei{ky} \me{\rr}{{\cal A}_k}{\rr'}\emi{ky} , \nn \me{\rr}{{\cal B}}{\rr'} &=& \frac{a}{2\pi}  \intk dk \; \ei{ky} \me{\rr}{{\cal B}_k}{\rr'}\emi{ky' } ,\eea where ${\cal A}_k$ and ${\cal B}_k$ are periodic in $\rr$ and $\rr'$ {\it separately}. The diagonal element of the product is then \begin{widetext}
\bea \me{\rr}{{\cal A B}}{\rr} &=& \frac{a^2}{(2\pi)^2}  \intk dk \intk dk' \ei{(k - k')y} \int_{-\infty}^\infty dx"  \int_{-\infty}^\infty  dy" \;  \me{\rr}{{\cal A}_k}{\rr"} \ei{(k'-k) y"} \me{\rr"}{{\cal B}_{k'}}{\rr} \nn &=& \frac{a}{2\pi}  \intk dk  \int_{-\infty}^\infty dx"  \int_0^a dy" \;  \me{\rr}{{\cal A}_k}{\rr"} \me{\rr"}{{\cal B}_{k}}{\rr} \nn &=& \frac{a}{2\pi}  \intk dk \;  \me{\rr}{{\cal A}_k {\cal B}_{k}}{\rr} .\eea 
\end{widetext}
This contraction is associative and can be repeated for three operators. It can also be generalized to system periodic in 2 or 3 dimensions, with an obvious change of notations.


\begin{thebibliography}{10}
\bibitem{herma}
{ We borrow the term "hermaphrodite'' from the nautical nomenclature:
  ``hermaphrodite brig'' is a two-masted ship rigged with a different kind of
  sails on each mast}.
\end{thebibliography}
\end{document}